\documentstyle[preprint,eqsecnum,aps,epsf,rotate,floats]{revtex}

\def\lqcd{\Lambda_{\rm QCD}}
\def\jp{J/\psi}
\def\psip#1{\psi_{\mathbf{#1}}}
\def\chip#1{\chi_{\mathbf{#1}}}
\def\bsigma{\mbox{\boldmath $\sigma$}}
 
\begin{document}

\preprint{
\vbox{\halign{&##\hfil    \cr
%	& hep-ph/0012062 \cr
	& CMU-0005 \cr
	&  FERMILAB-Pub-00/321-T \cr
	}}}
\title{Power Counting and Effective Field Theory for Charmonium}

\author{Sean Fleming and I.~Z.~Rothstein}
\address{Department of Physics, Carnegie Mellon University,
Pittsburgh, PA 15213, U.S.A}

\author{Adam K.~Leibovich}
\address{Theory Group, Fermilab, P.O. Box 500, Batavia, IL 60510, U.S.A.}

\maketitle

\begin{abstract}
\tighten{
\noindent\hspace*{-0.34cm} We hypothesize that the correct power
counting for charmonia is in the parameter $\lqcd/m_c$, but is not
based purely on dimensional analysis (as is HQET). This power counting
leads to predictions which differ from those resulting from the usual
velocity power counting rules of NRQCD. In particular, we show that
while $\lqcd/m_c$ power counting preserves the empirically verified
predictions of spin symmetry in decays, it also leads to new
predictions which include: A hierarchy between spin singlet and
triplet octet matrix elements in the $J/\psi$ system.  A quenching of the
net polarization in production at large transverse momentum.  No end
point enhancement in radiative decays.  We discuss explicit tests
which can differentiate between the traditional and new theories of
NRQCD.}
\end{abstract}

\vfill
\tighten{
\section{Introduction}

Quarkonia have proven to be fruitful in helping us gain a better
understanding of QCD. For large enough valence quark masses the system
should be dominated by Coulomb exchange in the perturbative regime.
Fortunately, the physical valence quark masses seem to be too small
for the states to be truly insensitive to non-perturbative effects,
and thus give a window on the more interesting aspects of QCD.  In
order to systematically study these effects we need to separate the
long distance from the short distance physics. This can be
accomplished by writing down a proper effective field theory to
describe the infra-red.  The theory should provide a power counting
which determines which operators are relevant. In most effective
theories this power counting is based upon dimensional
analysis. However, for non-relativistic QCD (NRQCD) \cite{BBL}
this is not the case. Instead, it is an expansion in the parameter
$v$, the relative velocity of the valence quarks. This power counting
presupposes that the states are Coulombic, at least to the extent that
$\alpha_s(mv)\simeq v$, and leads to the  result that operators
of the same dimension may be of different orders in the power
counting.  This methodology has been applied to the $J/\psi$ as well
as the $\Upsilon$ systems.  While it seems quite reasonable to apply
this power counting to the $\Upsilon$ system, it is not clear, as we
will discuss in more detail below, that it should apply for the
$J/\psi$ system. Indeed, we believe that the data is hinting
towards the possibility that a new power counting is called for in the
charmed system.

In Ref.~\cite{BBL}, the authors showed how to utilize NRQCD to predict
decay rates as well as production rates in a systematic double
expansion in $\alpha_s$ and $v$.  These predictions have met with
varying degrees of success. For instance, it is possible to explain
$J/\psi$ and $\psi'$ production at the Tevatron, though the initial
data on the polarization of these states at large transverse momentum
\cite{poldata} seems to be at odds with the NRQCD prediction. In
addition, there is an unexpected hierarchy of matrix elements in the
charmed system which does not seem to be there in the bottom
system. Furthermore, there is a discrepancy between theoretical
expectations and data for the end point spectrum of inclusive
radiative decays.  While we do not believe that any one of these
pieces of evidence, on their own,
 is strong enough to warrant introduction of a new theory, 
it seems  to us that the evidence, taken
as a whole, seems to be telling us that the effective field theory
which best describes the $J/\psi$ system may not be the same theory
which best describes the $\Upsilon$.  The purpose of this paper is to
present an alternative charmonium power counting, first discussed in
\cite{martin} and later utilized to study the quark-antiquark
potential \cite{Petal}, which leads to predictions which seem to have
better agreement with the data.

\section{background}
A general decay process may be written in factorized form \cite{BBL}
\begin{equation}
\Gamma_{J/\psi}=\sum  C_{^{2S+1} L_J}(m,\alpha_s)\langle \psi \mid
O^{(1,8)}(^{2S+1}L_J) \mid \psi \rangle.
\end{equation}
The matrix element represents the long distance part of the rate and
may be thought of as the probability of finding the heavy quarks in
the relative state $n$, while the coefficient
$C_{^{2S+1}L_J}(m,\alpha_s)$ is a short distance quantity calculable
in perturbation theory. The sum over operators may be truncated as an
expansion in the relative velocity $v$.

Similarly, production cross sections may be written as
\begin{equation}
\label{prodrate}
d\sigma =\sum_n d\sigma_{i+j\rightarrow Q\bar{Q}[n]+X}\langle 0 \mid
{\sc O}^H_n
\mid 0 \rangle.
\end{equation}
Here $d\sigma_{i+j\rightarrow Q\bar{Q}[n]+X}$ is the short distance
cross section for a reaction involving two partons, $i$ and $j$, in the
initial state, and two heavy quarks in a final state, labeled by $n$,
plus $X$.  This part of the process is calculable in perturbation
theory, modulo the possible structure functions in the initial
state. The production matrix elements, which differ from those used in
the decay processes, describe the probability of the short distance
pair in the state $n$ to hadronize, inclusively, into the state of
interest.  The relative size of the matrix elements in the sum are
again fixed by the power counting which we will discuss in more detail
below.

The formalism for decays is on the same footing as the operator
product expansion (OPE) for non-leptonic decays of heavy quarks, while
the production formalism assumes factorization, which is only proven,
and in some applications of production this is not even the case, in
perturbation theory \cite{factor}. The trustworthiness of
factorization depends upon the particular application, as we will
discuss further in the body of the paper.  We have reviewed these
results here to emphasize the point that when we test this theory we
are really testing both the factorization hypothesis as well the
validity of the effective theory as applied to the $J/\psi$ system.
Thus, we must be careful in assigning blame when we find that our
theory is not agreeing with the data.

While NRQCD has allowed for successful fits of the data (in particular
we have $J/\psi$ and $\psi^\prime$ production at the Tevatron in mind), its
predictive power has yet to stand any stringent test.\footnote{One
simple test, which has yet to be performed, is to compare the values
of the decay and production singlet matrix elements, which are
predicted to be equal at leading order in $v$ \cite{BBL}. 
To date the production singlet matrix elements have yet to
be extracted and compared to the decay singlet matrix elements. These
extractions can easily done using the direct $\jp$ production data at
CLEO \cite{IR}.}  Indeed, one robust prediction of the theory, namely
that production at large transverse momentum is almost purely
polarized \cite{CW95,BR96b,BL}, seems to be at odds with the initial
data.\footnote{The data still has rather large error bars, so we
should withhold judgment until the statistics improves.}  Other
predictions such as the ratio of $\chi_1/\chi_2$ in fixed target
experiments and the photon spectrum in inclusive radiative decays also
seem to disagree with the data, as we shall discuss in more detail
below. We are left with two obvious possibilities: 1) The power
counting of NRQCD does not apply to the $J/\psi$ system.  2)
Factorization is violated ``badly'', meaning that there are 
large power corrections.
The purpose of this paper is to explore the first possibility.

If we assume that NRQCD does not apply to the $J/\psi$ system, then we
must ask: is there another effective theory which does correctly
describe the $J/\psi$?  One good reason to believe that such a
theory does exist is that NRQCD, as formulated, does correctly predict
the ratios of decay amplitudes for exclusive radiative decays. Using
spin symmetry the authors of \cite{CW95} made the following
predictions:
\begin{eqnarray}
&&\Gamma(\chi_{c0}\rightarrow J/\psi+\gamma):
\Gamma(\chi_{c1}\rightarrow J/\psi+\gamma):\Gamma(\chi_{c2}\rightarrow
J/\psi+\gamma)
:\Gamma(h_{c}\rightarrow \eta_c+\gamma) \nonumber \\
&&\phantom{\qquad}=0.095:0.20:0.27:0.44\quad (theory)\nonumber \\
&&\phantom{\qquad}= 0.092\pm 0.041 :0.24\pm 0.04:0.27\pm
   0.03:unmeasured\quad(experiment).
\end{eqnarray}
Thus, we would like to find an alternative formulation (power
counting) of NRQCD which preserves these predictions yet yields
different predictions in other relevant processes.  Before discussing
this alternative power counting, we must briefly review the standard
formulation.

The rest of this paper is structured as follows.  We will first review
the standard power counting used for predictions to date. Then we will
offer a new power counting and discuss how the two theories differ in
their treatment of several relevant observables, as well as how the
theories fair against the data. We close with some remarks regarding
the validity of factorization in various observables.

\subsection{NRQCD power counting}

The power counting depends upon the relative size of the four scales
$(m,mv,mv^2,\lqcd)$. If we take $m>mv> mv^2\simeq \lqcd$, then the
bound state dynamics will be dominated by exchange of Coulombic gluons
with $(E\simeq mv^2,\vec{p}=m\vec{v})$.  This hierarchy has been
assumed in the NRQCD calculation of production and decay rates and is
most probably the reasonable choice for the $\Upsilon$ system, where
$mv \sim 1.5 {\rm\ GeV}$. However, whether or not it is correct for
the $\jp$, where $mv \sim 700 {\rm\ MeV}$ remains to be seen.

The power counting can be established in a myriad of different ways.
Here we will follow the construction of \cite{LMR}, which we now
briefly review. There are three relevant gluonic modes \cite{BSI}: the
Coulombic $(mv^2,mv)$, soft ($mv,mv$) and ultrasoft $(mv^2,mv^2)$. The
soft and Coulombic modes can be integrated out leaving only ultrasoft
propagating gluons. In the process of integrating out these modes we
must remove those large modes from the quark field. This is
accomplished by rescaling the heavy quark fields by a factor of
$\exp(i\vec{p}\cdot \vec{x})$ and labeling them by their three
momentum $\vec{p}$.  The ultrasoft gluon can only change residual
momenta and not labels on fields.  This is analogous to HQET, where
the four-velocity labels the fields and the non-perturbative gluons
only change the residual momenta \cite{MWbook}.  This rescaling must
also be done for soft gluon fields \cite{HG} which, while they cannot
show up in external states, do show up in the
Lagrangian.\footnote{Thus the nomenclature is slightly misleading
since we have not removed these fields from the Lagrangian.}  After
this rescaling a matching calculation leads to the following tree
level Lagrangian \cite{LMR}
\begin{eqnarray}\label{nrqcd:1}
{\cal L} &=&  
  \sum_{\mathbf p}
 \psip p ^\dagger   \Biggl\{ i D^0 - {\left({\bf p} 
 \right)^2 \over 2 m} \Biggr\} 
 \psip p   - 4 \pi \alpha_s
\sum_{q,q^\prime\mathbf p,p^\prime}\Biggl\{{1\over q^0} 
 \psip {p^\prime} ^\dagger  \left[A^0_{q^\prime},A^0_q \right]
  \psip p \nonumber \\
&&  + {g^{\nu 0} \left(q^\prime-p+p^\prime\right)^\mu -
  g^{\mu 0} \left(q-p+p^\prime\right)^\nu + g^{\mu\nu}\left(q-q^\prime \right)^0
  \over \mathbf \left( p^\prime-p \right)^2}
  \psip {p^\prime} ^\dagger  \left[A^\nu_{q^\prime},A^\mu_q \right]
  \psip p \Biggr\}\nonumber \\
 &&\qquad\qquad + \psi \leftrightarrow \chi,\ T \leftrightarrow \bar T + \sum_{\mathbf p,q}
  {4 \pi \alpha_s  \over \mathbf \left( p-q \right)^2} 
  \psip q ^\dagger T^A \psip p \chip {-q}^\dagger  \bar T^A \chip {-p} + \ldots
\end{eqnarray}
where we have retained the lowest order terms in each sector of the
theory. The matrices $T^A$ and $\bar T^A$ are the color matrices for
the $\bf 3$ and $\bf \bar 3$ representations, respectively.  Notice
that the kinetic piece of the quark Lagrangian is just described by a
label. This is a result of the dipole expansion \cite{GR} which is
used to get a homogeneous power counting.  The last term is the
Coulomb potential, which is leading order and must be resummed in the
four-quark sector, while the other non-local interactions arise from
soft gluon scattering.

Now all the operators in the Lagrangian have a definite scaling in
$v$.  The spin symmetry, which will play such a crucial role in the
polarization predictions, is manifest. The two subleading interactions
which will dominate our discussion are the ``electric dipole'' ($E1$)
\begin{equation}
{{\cal L}_{E1}}=\psip p  ^\dagger \frac{\vec{p}}{m}\cdot \vec{A} \psip p,
\end{equation}  
and ``magnetic dipole'' ($M1$)
\begin{equation}
{{\cal L}_{M1}}= c_F\, g \psip p ^\dagger {{\bf \bsigma \cdot B} \over 2
m}\psip p.
\end{equation} 
The $E1$ interaction is down by a factor of $v$ while the $M1$ is down
by a factor of $v^2$. The extra factor of $v$ stems from the fact that
the magnetic gluons are ultrasoft,\footnote{One may wonder why the
emission of a soft gluon cannot lead to the enhancement of the
magnetic transition operator. However, the emission of such a gluon
leaves the quark off-shell and contributes a pure counter-term to the
matching \cite{martin}.} and the derivative operator therefore picks up
a factor of $v^2$.  These operators play a crucial role in the so-called
octet mechanism.

\subsection{New Power Counting}
Let us now consider the alternate hierarchy $m>mv\sim \lqcd$.  One
might be tempted to believe that in this case the power counting
should be along the lines of HQET, where the typical energy and
momentum exchanged between the heavy quarks is of order $\lqcd$.
However, this leads to an effective theory which does not correctly
reproduce the infra-red physics. With this power counting, the leading
order Lagrangian would simply be
\begin{equation}
{\cal L_{\rm HQET}}=\psi_v ^\dagger D_0 \psi_v,
\end{equation}
where the fields are now labeled by their four velocity.  This is a
just a theory of time-like Wilson lines (static quarks) which does not
produce any bound state dynamics.  Thus we are forced to the
conclusion that the typical momentum is of order $\lqcd$, whereas the
typical energy is $\lqcd^2/m$.  The dynamical gluons are now all of
the type $(\lqcd,\lqcd)$, as the on-shell ultrasoft modes get cut-off by the
confinement scale.  Therefore, one no longer labels fields by their three
velocities.  The only label is the four velocity of the heavy
quark. However, the $D^2/(2m)$ is still relevant and their is no
dipole expansion.  We can not resist the temptation of 
introducing yet another
acronym,\footnote{We stole this bit of prose from \cite{BFL*}.} and
call this theory NRQCD$_c$, while we will refer to the traditional
power counting as NRQCD$_b$ as we assume that it does describe the
bottom system.\footnote{In the language of \cite{pNRQCD}, NRQCD$_b$
would correspond to pNRQCD and NRQCD$_c$ would correspond to NRQCD. We
chose to introduce these new acronyms because calling NRQCD$_c$ NRQCD
would be misleading, since the original NRQCD, as defined in
\cite{BBL}, is indeed distinct from NRQCD$_c$.  We thus believe that
our labeling will be the simplest for our purposes and hope the
community will indulge us in our, what may be perceived as gratuitous,
acronymization.}
 
The power counting of this theory is now along the lines of HQET where
the expansion parameter is $\lqcd/m_Q$.  However the residual energy of
the quarks is order $\lqcd^2/m_Q$, while the residual three momentum
is $\lqcd$.  Thus one must be careful in the power counting to
differentiate between time and spatial derivatives acting on the quark
fields.  As far as the phenomenology is concerned, perhaps the most
important distinction between the power counting in NRQCD$_c$ and
NRQCD$_b$ is that the magnetic and electric gluon transitions are now
of the same order in NRQCD$_c$.  This difference in scaling does not
disturb the successes of the standard NRQCD$_b$ formulation but does
seem help in some of its shortcomings.

\section{Lifetimes}

In the case of inclusive decays the use of effective field theory put
theoretical calculations on surer footing. Previous to the advent of
NRQCD, inclusive decays were written as a product of a short distance
decay amplitude and a long distance wave function which was usually
taken from potential models \cite{Buch}
\begin{equation}
\Gamma_{J/\psi}=\mid \psi(0)\mid^2 C(m,\alpha_s).
\end{equation}
Most of the time this formalism is adequate, however there is the
question of the scheme dependence of the potential wave function
beyond leading order. Beyond this drawback is the question of how to
factor infra-red divergences in $P$ wave decays.  Within the
effective field theory approach, however, these issues are
clarified. The rate is now written as
\begin{equation}
\Gamma_{J/\psi}=\sum  C_{^{2S+1} L_J}(m,\alpha_s)\langle \psi \mid
O_{(1,8)}(^{2S+1}L_J) \mid \psi \rangle.
\end{equation}
The operator matrix element gives the probability to find the quarks
within the hadron in the state $^{2S+1} L_J$. The quarks can be either
in a relative singlet or octet state, hence the subscript $(1,8)$. The
matrix elements are well defined scheme dependent quantities, which
can be measured on the lattice, or extracted from the data
\cite{bodwin}, and have definite scalings in $v$. For instance,
consider the operator $\langle \chi_J \mid O_1(^{3}P_J) \mid \chi_J
\rangle$. We would expect this operator to dominate the decay of
$\chi$ states, given that the quantum number of the short distance
quark pair match the quantum number of the final state. However, this
is not the case \cite{pwave}.  The operator $\langle \chi_J \mid
O_8(^{3}S_1) \mid \chi_J \rangle$ is of the same order. This can be
seen from the fact that the $P$ wave operator comes with two spatial
derivatives.  The octet $S$ operator vanishes at leading order, since
there is no $S$ wave component in the leading order hadronic state in
the effective theory.  The first non-vanishing contribution comes from
two insertion of $E1$ operators into time ordered products with
$O_8(^{3}S_1)$. Thus both the singlet $P$ wave operator and the octet
$S$ wave operator scale as $v^2$. Furthermore, the inclusion of this
operator into the rate allows for the proper absorption of infrared
divergences in the $P$ wave decays into octet $S$ wave matrix
elements. This should be considered a formal success of the effective
field theory.  Any change in the power counting will not change this
success, as the scaling of an operator is independent of its
renormalization group properties. Such a change could only effect the
relevance of the infra-red divergence, in a technical sense.

The advent of NRQCD had little impact on the phenomenology of
inclusive decays because it simply justified previous calculations of
the total width. However,  one novel prediction of NRQCD was found in the
end point spectrum of inclusive radiative decays \cite{RW,MW}.
Radiative decays, as opposed to hadronic decays,\footnote{We are
ignoring photon fragmentation for the moment.} have the advantage that
they are subject to an operator product expansion, thus rely less upon
local-duality assumptions.  The integration over the photon energy
smears through resonances and thus one may expect the prediction
to be more trustworthy.  We may reliably calculate the photon spectrum
itself if we smear over regions of phase space which are larger than
$\lqcd$ \cite{RW}.

In NRQCD, the decay of the $J/\psi$ is dominated by the $^3S_1$
singlet operator with the octet operators being suppressed by
$v^4\pi/\alpha_s$.  However, due to the singular nature of the octet
Wilson coefficients (it is a delta function at leading order in
$\alpha_s$) they can become leading order near the end point of the
spectrum.  Of course we do not expect a delta function spike at the
end point since the spike should be smeared out due to bound state
dynamics, among other effects \cite{wolf}, which can be taken into
account by the introduction of structure functions.

In \cite{RW} it was shown that if one smeared the photon spectrum over
a range of order $mv^2$, then the spectrum would receive a leading
order corrections from the octet matrix elements which are peaked at
the end point.  In the standard hierarchy such a smearing is
satisfactory since it corresponds to smearing over $mv$ in hadronic
mass which is larger than $\lqcd$. However, in the new power counting
this is no longer true and, given that the OPE breaks down in the
region where the octet was suppose to dominate, it is no longer true
that we can predict any peak with reliability.  If we now consider the
data, we see that for the $J/\psi$ the data is monotonically
decreasing \cite{Jdata}. On the other hand the $\Upsilon$ data does
show a bump out at larger values of the photon energy. We wish to take
this as support for the new power counting in the $J/\psi$ system. But
we must be careful since there are other effects which become
important at the end point which we have not taken into account.  For
instance, near the end point there are large radiative corrections
which are known to resum into a Sudakov suppression.  However, we
would not expect this effect to completely eliminate the bump, just
cut it off at larger energies. Nonetheless a complete calculation of
the resummed Sudakov effects in the end point spectrum of $\Upsilon$
decays is needed.

\section{Hadro-Production}
As discussed in section II a general production process may be
written in the factorized form (\ref{prodrate}).  The long distance
part of the process involves the hadronization of the heavy quarks in
the state $n$ into the hadron of choice $H$.  The matrix element in
Eq.~(\ref{prodrate}) is written as
\begin{eqnarray}
\langle 0 \mid O^H_n \mid 0 \rangle &=&  
 \langle 0 \mid \psi^\dagger \Gamma^{n^\prime} \chi \mid\sum_X
H+X\rangle  \langle H+X \mid \chi^\dagger \Gamma^n \psi \mid
 0\rangle\\
&\equiv& \langle O^H_n \rangle.\nonumber
\end{eqnarray}  
The tensor $\Gamma^n$ operates in color as well as spin space and also
contains possible derivatives. This tensor determines the order of the
matrix element. If the quantum numbers $n$ do not match the quantum
numbers of the hadron, then the matrix element vanishes, as the
hadronic states are those of the effective field theory, and are pure
in the sense of a Fock space expansion. To get a non-vanishing result
one must insert subleading operators into a time ordered product with
the operator $O^H_n$.  The number and order of the inserted operators
determine the scaling of the matrix element, as we detail below.

\subsection{Collider experiments}
The leading order contribution to $\psi$ production in the original
$v$ power counting scheme is through the color-singlet matrix element
$\langle O_1^\psi(^3S_1)\rangle$, since the quantum numbers of the
short distance quark pair matches those of the final state.  All other
matrix elements need the insertion of operators into time ordered
products to give a non-zero result.  Unlike the case of the $\chi$
discussed earlier, all other matrix elements are suppressed compared
to the color-singlet matrix element above.  For instance, the matrix
element $\langle O_8^\psi(^1S_0)\rangle$ vanishes at leading order.
The first non-vanishing contribution comes from the insertion of two
$M1$ operators into time ordered products, thus giving a $v^4$
suppression.  The scalings of the relevant matrix elements for $\psi$
production in NRQCD$_b$ are shown in Table~\ref{psiMEscaling}.  It
appears from just the $v$ counting that only the color-singlet
contribution is important.  But the other contributions can be
enhanced for kinematic reasons.  At large transverse momentum,
fragmentation type production dominates \cite{BY}, and only the
$\langle O_8^\psi(^3S_1)\rangle$ contribution is important.  Without
the color-octet contributions ({\it i.e.}, the Color-Singlet Model),
the theory is below experiment by about a factor of 30.  By adding the
color-octet contribution the fit to the data is very good
\cite{BF}.  The new power counting must also reproduce this success.
\begin {table}[t]
\begin {center}
\begin {tabular}{c|cccc}
\\
 & $\langle O_1^\psi(^3S_1)\rangle$  
 & $\langle O_8^\psi(^3S_1)\rangle$ 
 & $\langle O_8^\psi(^1S_0)\rangle$ 
 & $\langle O_8^\psi(^3P_0)\rangle$ \\\\
\hline
\\
NRQCD$_b$  & $v^0$ &  $v^4$ &  $v^4$ &  $v^4$ \\\\
\hline
\\
NRQCD$_c$  & $(\lqcd/m_c)^0$ 
        & $(\lqcd/m_c)^4$ 
        & $(\lqcd/m_c)^2$ 
        & $(\lqcd/m_c)^4$ \\\\
\end {tabular}
\end {center}
\caption{Scaling of matrix elements relevant for $\psi$ production in
NRQCD$_b$ and NRQCD$_c$.}
\label{psiMEscaling}
\end {table}

The relative size of the different matrix elements change in
NRQCD$_c$.  In particular, the $M1$ transition is now the same order
as the $E1$ transition.  The new scalings are shown in
Table~\ref{psiMEscaling} \cite{tab}.  Due to the dominance of
fragmentation at large transverse momentum, we need to include effects
up to order $(\lqcd/m_c)^4$, since the $\langle
O_8^\psi(^3S_1)\rangle$ matrix element will still dominate at large
$p_T$.

Is this consistent?  The size of the matrix elements is a clue.
Extraction of the matrix elements uses power counting to limit the
number of channels to include in the fits.  Calculating $J/\psi$ and
$\psi'$ production up to order $(\lqcd/m_c)^4$ in NRQCD$_c$ requires
keeping the same matrix elements as in NRQCD$_b$.  Previous
extractions of the matrix elements only involve the linear combination
\begin{equation}
M_r^\psi = \langle O_8^\psi(^1S_0)\rangle 
   + \frac{r}{m_c^2}\langle O_8^\psi(^3P_J)\rangle, 
\end{equation}
with $r\approx3-3.5$, since the short-distance rates have similar size
and shape.  In the new power-counting, we can just drop the
contribution from $\langle O_8^\psi(^3P_J)\rangle$, since it is down
by $(\lqcd/m_c)^2\sim 1/10$ compared to $\langle
O_8^\psi(^1S_0)\rangle$.  It is the same order as $\langle
O_8^\psi(^3S_1)\rangle$, but is not kinematically enhanced by
fragmentation effects.  The extraction from \cite{BKL00} would then
give for the $J/\psi$ and $\psi'$ matrix elements
\begin{eqnarray}
\langle O_8^{\jp}(^1S_0)\rangle:\langle O_8^{\jp}(^3S_1)\rangle
   &=& (6.6 \pm 0.7) \times 10^{-2} :(3.9 \pm 0.7) \times 10^{-3}
   \approx 17 : 1 \,, \nonumber\\
\langle O_8^{\psi'}(^1S_0)\rangle:\langle O_8^{\psi'}(^3S_1)\rangle
   &=& (7.8 \pm 3.6) \times 10^{-3} :(3.7 \pm 0.9) \times 10^{-3}
   \approx 2 : 1.
\end{eqnarray}
Other extractions have various values of the hierarchy, ranging from
$3:1$ to $20:1$ \cite{hier}.  While the relation of the color-octet
matrix elements in the $\jp$ system is indeed in agreement with the
NRQCD$_c$ power counting, the $\psi'$ does not look to be
hierarchical. However, it should be noted that the statistical errors
in the $\psi^\prime$ extraction, quoted above, are quite large.
Furthermore, there are also large uncertainties introduced in the
parton distribution function.  The above ratios used the CTEQ5L parton
distribution functions.  If we take the central values from
\cite{BKL00} for the MRST98LO distribution functions, we find the
ratio $3:1$. On the other hand, the $J/\psi$ extraction is much less
sensitive to the choice of distribution function.  Given the
statistical and theoretical errors, it clear that the $\psi^\prime$
ratio is not terribly illuminating.

Let us now consider the  extraction of these color-octet matrix
elements in the $\Upsilon$ sector~\cite{BFL}, where according to
NRQCD$_b$ power counting there is should be no hierarchy:
\begin{eqnarray}
\langle O_8^{\Upsilon(3S)}(^1S_0)\rangle:
\langle O_8^{\Upsilon(3S)}(^3S_1)\rangle &=& 
(5.4 \pm 4.3^{+3.1}_{-2.2}) \times 10^{-2} :
(3.6 \pm 1.9^{+1.8}_{-1.3}) \times 10^{-2}
\approx 1 : 1, \\
\langle O_8^{\Upsilon(2S)}(^1S_0)\rangle:
\langle O_8^{\Upsilon(2S)}(^3S_1)\rangle &=& 
(-10.8 \pm 9.7^{-3.4}_{+2.0}) \times 10^{-2} :
(16.4 \pm 5.7^{+7.1}_{-5.1}) \times 10^{-2}
\approx 1 : 1 \nonumber, \\
\langle O_8^{\Upsilon(1S)}(^1S_0)\rangle:
\langle O_8^{\Upsilon(1S)}(^3S_1)\rangle &=& 
(13.6 \pm 6.8^{+10.8}_{-7.5}) \times 10^{-2} :
(2.0 \pm 4.1^{-0.6}_{+0.5}) \times 10^{-2}
\approx 6 : 1. \nonumber 
\end{eqnarray}
For the $\Upsilon(3S)$ and $\Upsilon(2S)$ we observe that there is
indeed no hierarchy, while for the $\Upsilon(1S)$ it appears like
there may be a hierarchy \cite{BFL}. However, it is not possible to
draw any strong conclusions from these data because the errors on the
extractions are large. In fact the ratio for the $\Upsilon(1S)$
color-octet matrix elements is $1:1$ within the one sigma
errors. Furthermore, these matrix elements are those extracted
subtracting out the feed down from the higher states. Thus, the
extraction of the $\Upsilon(1S)$ matrix elements actually have larger
errors since the errors accumulate when we make the subtractions of
the feed down components. Finally, we should add that, while
phenomenologically it is perfectly reasonable to define the subtracted
matrix elements, we believe that, since the matrix elements are
inclusive, one should not subtract out the feed down from hadronic
decays when checking the power counting. In principle this subtraction
should not change things by orders of magnitude, but nonetheless can
have a significant effect. Indeed, if one compares the ratios for
inclusive matrix elements, which do not have the accumulated error,
then the ratios come out to be $1:1$, even for the $\Upsilon(1S)$
\cite{BFL}.\footnote{The authors of \cite{DS} have values of the
extracted matrix elements that differ from \cite{BFL}.  They use
PYTHIA to model initial state gluon radiation, and use data at small
values of $p_T$ in the extraction.  Since we are worried about
breakdown of factorization, we prefer to restrict our analysis to data
points where factorization should hold.}

\subsection{Fixed-target experiments}
There are several phenomenological differences between NRQCD$_c$ and
NRQCD$_b$ in fixed target experiments \cite{BR96,GS}.  Here we will
focus on $\psi$ production and the predicted ratio of production cross
sections $\sigma(\chi_1)/\sigma(\chi_2)$ in NRQCD$_c$.

\begin {table}[t]
\begin {center}
\begin {tabular}{c|ccccc}
\\
 & $\langle O_1^\chi(^3P_J)\rangle$  
 & $\langle O_8^\chi(^3S_1)\rangle$ 
 & $\langle O_8^\chi(^1S_0)\rangle$ 
 & $\langle O_8^\chi(^3P_J)\rangle$
 & $\langle O_8^\chi(^1P_1)\rangle$ \\\\
\hline
\\
NRQCD$_b$  & $v^2$ &  $v^2$ &  $v^6$ &  $v^6$ & $v^6$ \\\\
\hline
\\
NRQCD$_c$  & $(\lqcd/m_c)^2$ 
        & $(\lqcd/m_c)^2$ 
        & $(\lqcd/m_c)^4$ 
        & $(\lqcd/m_c)^6$ 
	& $(\lqcd/m_c)^4$\\\\
\end {tabular}
\end {center}
\caption{Scaling of matrix elements relevant for $\chi$ production in
NRQCD$_b$ and NRQCD$_c$.}
\label{chiMEscaling}
\end {table}

At order $\alpha_s^2$, $\psi$s are produced via quark-antiquark fusion
through the $\langle O_8^\psi(^3S_1)\rangle$ matrix element and
through gluon fusion through the $\langle O_8^\psi(^1S_0)\rangle$ and
$\langle O_8^\psi(^3P_0)\rangle$ matrix elements, in the linear
combination $M_7^\psi$.  At fixed-target energies, the contribution to
$\psi$ production from $\langle O_8^\psi(^3S_1)\rangle$ is numerically
irrelevant because gluon fusion dominates.  The difference between the
NRQCD$_c$ prediction and the NRQCD$_b$ analysis done in \cite{BR96}
lies in the expected size of the matrix elements in $M_7^\psi$ (called
$\Delta_8(\psi)$ in \cite{BR96}), since in NRQCD$_c$ the $^3P_0$
matrix element is down by $(\lqcd/m_c)^2$.  However, since the $^3P_0$
matrix element is enhanced by a factor of $7$, it is important to keep
this formally subleading contribution.  Furthermore, there is a very
large scale and PDF dependence in these extractions, so it is not clear
whether or not we can learn anything from comparisons with the
Tevatron extractions.

The $\chi_1/\chi_2$ production ratio has the nice property that it is
relatively insensitive to the charmed quark mass \cite{BR96}.
$\chi_1$ production is suppressed as it can not be produced at leading
order in the singlet channel. The formally leading order $\chi_1$
channel is $^3S_1^{(8)}$ through quark-antiquark fusion.  In both
NRQCD$_b$ and NRQCD$_c$ this formally leading order contribution is
actually smaller than the subleading contributions coming from other
octet operators ($\langle O_8^{\chi_1}(^3P_0)\rangle$ and $\langle
O_8^{\chi_1}(^3P_2)\rangle$ in NRQCD$_b$, and $\langle
O_8^{\chi_1}(^1S_0)\rangle$ in NRQCD$_c$) due to the fact that these
other channels are initiated by gluon-gluon fusion.  The scalings for
the $\chi$ matrix elements are shown in Table~\ref{chiMEscaling}.  If
we ignore the quark initiated process then due to simplicity of the
$2\rightarrow 1 $ kinematics we may write the NRQCD$_b$ prediction as
\begin{equation}
\sigma_{\chi_1}/ \sigma_{\chi_2} \simeq \frac{75}{32}\frac{
\langle O_8^{\chi_1}(^1S_0)\rangle m^2_c +
3 \langle O_8^{\chi_1}(^3P_0)\rangle +
4 O_8^{\chi_1}(^3P_2)\rangle }
{\langle O_1^{\chi_2}(^3P_2)\rangle} \,,
\end{equation}
where numerically small contributions have been dropped.
This ratio is approximately $1/3$ if we take take $v^2\approx 0.3$.
Also, the ratio is independent of the center of mass energy, which
agrees with the data within errors \cite{kor}.

In NRQCD$_c$ we have (again neglecting the numerically small 
quark-antiquark initiated processes)
\begin{equation}
\sigma_{\chi1}/ \sigma_{\chi_2} \simeq \frac{75}{32}
\frac{\langle O_8^{\chi_1}(^1S_0)\rangle m^2_c} 
     {\langle O_1^{\chi_2}(^3P_2)\rangle} \,,
\end{equation}
where, once again, numerically small contributions have been dropped.
If we take $\lqcd/m_c\approx 1/3$ we get approximately the same
result. This estimate is so crude that it is not clear whether any
information can be gleaned from it. However, it does seem that in
either description the data \cite{c1c2data} is, on the average, larger
than these naive predictions.\footnote{One robust prediction, however
is that the ratio should be independent of $s$, which does seem to
agree with the data.}  This could very well be due to large
non-factorizable contributions, which we may expect to be enhanced in
NRQCD$_c$ (see the conclusions).

\section{Polarization}
$J/\psi$ and $\psi'$ are predicted to be transversely polarized at
large $p_T$ in NRQCD$_b$.  At large transverse momentum, the dominant
production mechanism is through fragmentation from a nearly on shell
gluon to the octet $^3S_1$ state.  The quark pair inherits the
polarization of the fragmenting gluon, and is thus transversely
polarized \cite{CW95}.  In NRQCD$_b$ the leading order transition to
the final state goes via two $E1$, spin preserving, gluon emissions.
Higher order perturbative fragmentation contributions \cite{BR96b},
fusion diagrams \cite{BK97,L97}, and feed-down for the $J/\psi$
\cite{BKL00} dilute the polarization some, but the prediction still
holds that as $p_T$ increases so should the transverse polarization.
Indeed, for the $\psi^\prime$, at large $p_T\gg m_c$, we expect nearly
pure transverse polarization.

The polarization of $J/\psi$ and $\psi'$ at the Tevatron has recently
been measured \cite{poldata} with large error bars.  The experimental
results show no or a slight longitudinal polarization, as $p_T$
increases. If, after the statistics improve, this trend continues,
then it will be the smoking gun that leads us to conclude that
NRQCD$_b$ is not the correct effective field theory for charmonia.

With NRQCD$_c$, the intermediate color-octet $^3S_1$ states hadronize
through the emission of either two $E1$ or $M1$ dipole gluons, at the
same order in $1/m_c$.  Since the magnetic gluons do not preserve
spin, the polarization of $\psi$ produced through the $\langle
O_8^{\psi}(^3S_1)\rangle$ can be greatly diluted.  The net
polarization will depend on the ratio of matrix elements
\begin{eqnarray}
R_{M/E}&&=
\nonumber \\
&&\frac{\int \prod_{\ell} d^4x_{\ell}
 \langle 0 \mid T(M_1(x_1)M_1(x_2)\psi^\dagger T^a\sigma_i \chi)
 \,a_H^\dag\,a_H\,T(M_1(x_3)M_1(x_4)\chi^\dagger T^a\sigma_i \psi) 
 \mid 0\rangle}
{\int \prod_{\ell}d^4x_{\ell}  
 \langle 0 \mid T(E_1(x_1)E_1(x_2)\psi^\dagger T^a\sigma_i \chi)
 \,a_H^\dag\,a_H\,T(E_1(x_3)E_1(x_4)\chi^\dagger T^a\sigma_i \psi) 
 \mid 0\rangle}
\end{eqnarray}
where
\begin{equation}
a_H^\dag\,a_H  = \sum_X\mid H+X\rangle \langle H+X \mid.
\end{equation}
The leads to the polarization leveling off at large $p_T$ at some
value which is fixed by $R_{M/E}$.  In Fig.~\ref{polar}, we show the
prediction for $J/\psi$ and $\psi'$ polarization at the Tevatron.  The
data is from \cite{poldata}.  The three lines correspond to
different values for $R_{M/E}$=(0 (dashed), 1 (dotted), $\infty$
(solid)). The dashed line is also the prediction for NRQCD$_b$. The
residual transverse polarization for $J/\psi$ at asymptotically large
$p_T$ is due to feed down from $\chi$ states.  The non-perturbative
corrections to our predictions are suppressed by $\lqcd^4/m^4$.

\begin{figure}[ht]
\centerline{\epsfxsize 14cm \epsffile{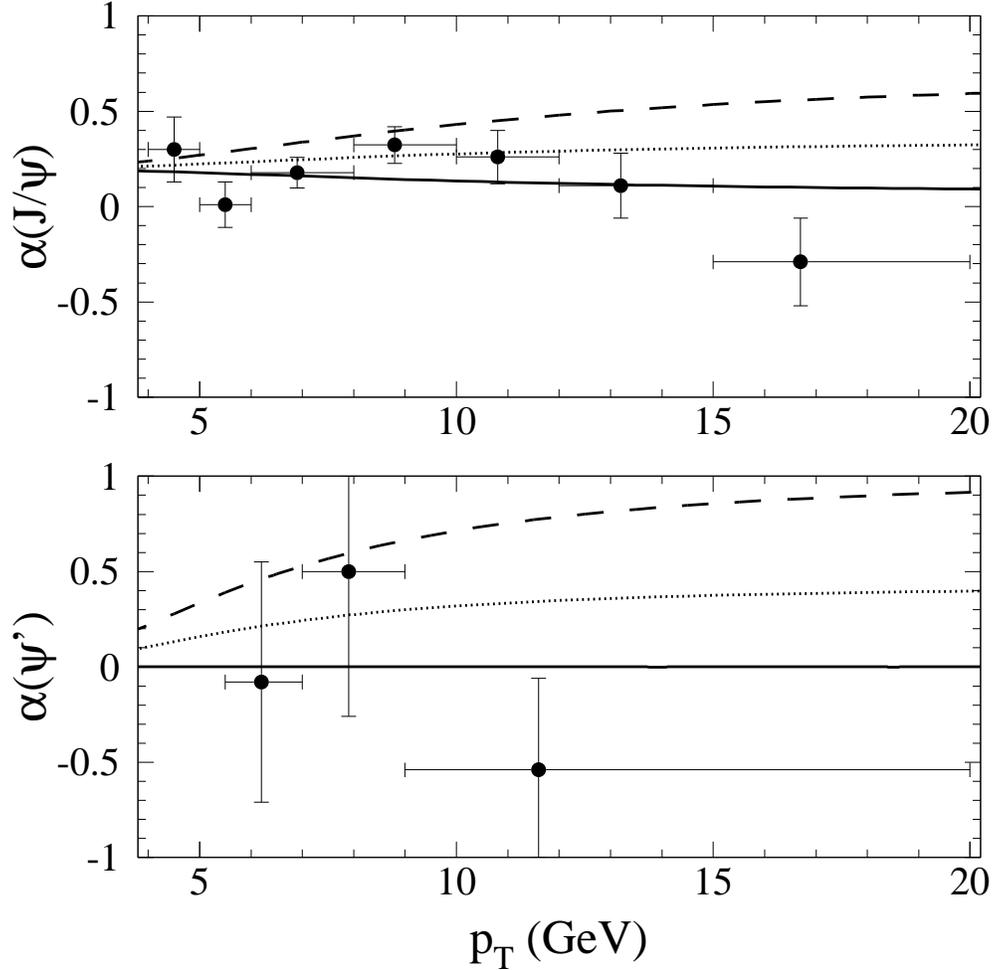}}
\caption{Predicted polarization in NRQCD$_c$ for $J/\psi$ and $\psi'$
at the Tevatron as a function of $p_T$.  The three lines correspond to
$R_{M/E}$=(0 (dashed), 1 (dotted), $\infty$ (solid)).  The dashed line
is also the prediction for NRQCD$_b$.}
\label{polar}
\end{figure}

\section{$B$ decays}
Another useful observable for differentiating between NRQCD$_c$ and
NRQCD$_b$ is charmonia production in $B$ decays.  Assuming perturbative
factorization, the $\psi$ production rate from semi-inclusive $B$
decays may be written as
\begin{equation}
\label{factform}
\Gamma(B\to H+X) = \sum_n C(b\to c\bar{c}[n]+x)\,\langle O^H_n\rangle.
\end{equation}
This expression is valid up to power corrections of order
$\lqcd/m_{b,c}$, which parameterize the non-factorizable
contributions.  (To this accuracy it is justified to treat the $B$
meson as a free $b$ quark.)  The short distance coefficients are
determined by the $\Delta B = 1$ effective weak Hamiltonian
\begin{equation}  
H_{eff} = \frac{G_{F}}{\sqrt{2}} \sum_{q=s,d} 
\left\{V_{cb}^\ast V_{cq} \,
\left[ \frac{1}{3} C_{[1]}(\mu) {\cal O}_1(\mu) + C_{[8]}(\mu) 
{\cal O}_8(\mu) \right] - V_{tb}^\ast V_{tq} \sum_{i=3}^6 
C_i(\mu) {\cal O}_i(\mu)\right\}
\label{eq:Heff}  
\end{equation}
containing the `current-current' operators   
\begin{eqnarray}
{\cal O}_1 &=&   
[\bar{c} \gamma_{\mu} (1-\gamma_5) c] \, 
[\bar{b} \gamma^{\mu} (1-\gamma_5) q],
\label{eq:opsing} \\  
{\cal O}_8 &=&    
[\bar{c}\,T^A \gamma_{\mu} (1-\gamma_5) c]\,  
[\bar{b}\,T^A \gamma^{\mu} (1-\gamma_5) q],
\label{eq:opoct}  
\end{eqnarray}
and the QCD penguin operators ${\cal O}_{3-6}$\ (precise definitions
may be found in the review \cite{buras}).  For the decays $B\to
\mbox{charmonium}\,+X$ it is convenient to choose a Fierz version of
the current-current operators such that the $c\bar{c}$ pair at the
weak decay vertex is either in a color-singlet or a color-octet
state. The coefficient functions are related to the usual $C_\pm$ by
\begin{eqnarray}
C_{[1]}(\mu) &=& 2 C_+(\mu) - C_-(\mu),  
\label{eq:c1nlo} \\  
C_{[8]}(\mu) &=&   C_+(\mu) + C_-(\mu).  
\label{eq:c8nlo}  
\end{eqnarray} 
In NRQCD$_b$ a naive power counting leads to the conclusion that the
leading order result is fixed by the $\langle O_1(^3S_1)\rangle$ operator
as all octet operators are suppressed by $v^4$. However, as pointed
out in Ref.~\cite{KLS96}, the fact that the Wilson coefficients
evaluated at the low energy scale are numerically hierarchical,
$C_1^2/C_8^2\approx 15$, actually leads to octet domination.  In
NRQCD$_b$ one would then get leading order contributions from all the
octet matrix elements, $\langle O^{\jp}_8(^3S_1)\rangle$, $\langle
O^{\jp}_8(^1S_0)\rangle$ and $\langle O^{\jp}_8(^3P_0)\rangle$, where
the contribution from the other $^3P_J$ states have been written in
terms of the $^3P_0$ contribution using spin symmetry.  

In NRQCD$_c$, the leading order octet contribution comes solely from the
$^1S_0$ operator which is suppressed by $\lqcd^2/m_c^2$.  Thus, we may
get a direct extraction of this matrix element from the decay
rate. However, at leading order in $\alpha_s$ the color-singlet
contribution is highly scale dependent. This is due to large scale
dependence in the value of $C_1(\mu)$. The authors of \cite{BR96}
found that the leading order singlet contribution varies by a factor
of ten as $\mu$ is varied between 2.5 and 10 GeV. This scale
dependence can be drastically reduced by working at next-to-leading
order (NLO) and using a combined expansion in $\alpha_s$ and the ratio
$C_1/C_8$ \cite{BE94}.  Using this expansion a NLO order calculation
found that within NRQCD$_b$ power counting one could extract the linear
combination \cite{BMR}
\begin{equation}
\label{central}
M_{3.1}^\psi({}^1\!S_0^{(8)},{}^3\!P_J^{(8)})
=\left\{\begin{array}{c}
\,\,\,1.5\cdot 10^{-2}\,\mbox{GeV}^3\qquad (J/\psi) \\
0.6\cdot 10^{-2}\,\mbox{GeV}^3\qquad (\psi').
\end{array}
\right.
\end{equation} 
In NRQCD$_c$ the result is all spin singlet and we would thus conclude
that $\langle O^{\jp}_8({}^1S_0)\rangle=1.5\cdot 10^{-2}$, which is
quite a bit smaller than the Tevatron extraction .

We may also consider how the new power counting effects the prediction
for the polarization of the $\psi$.  In NRQCD$_b$ the prediction for
the polarization, at leading order in $\alpha_s$, was given in
\cite{FHMN}. The angular distribution in the leptonic $J/\psi$ decay
may be written as
\begin{equation}
\frac{d\Gamma}{d\cos\theta}(\psi\rightarrow \mu^+\mu^-)(\theta)
  \propto 1+\alpha\, \cos^2\theta,
\end{equation}
where the angle $\theta$ is defined in the $\jp$ rest frame for which
the $z$-axis is aligned with direction of motion of the $\jp$ and
\begin{equation}
\alpha=\frac{\sigma(+)+\sigma(-)-2 \sigma(0)}{\sigma(+)+\sigma(-)+2 \sigma(0)}.
\end{equation}
Within NRQCD$_b$ the authors found that $\alpha$ lies within the range
$-0.4<\alpha<-0.1$ if the bottom quark mass lies between $4.4$ and
$5.0$ GeV and the octet matrix elements are allowed to vary within
their errors. This rather crude leading order prediction should be
reasonable as long as the scale dependent singlet piece is not
dominant, which it is not.

In NRQCD$_c$, given the color-octet $^1S_0$ dominance, we would
expect a quenching of the polarization, since, as discussed in the
case of hadro-production, the spin flipping hadronic transition
involved in the matrix element obeys helicial democracy. Using the
results of \cite{FHMN} we may write down the order $\alpha_s$
NRQCD$_c$ prediction for $\alpha$
\begin{equation}
\alpha=\frac{-0.39 \langle O^{\jp}_1(^3S_1)\rangle}
{ \langle O^{\jp}_1(^3S_1)\rangle+61 \langle O^{\jp}_8(^1S_0)\rangle},
\end{equation}
where we have kept the formally subleading $^1S_0$ contribution in the
denominator because of its large coefficient.  Since the singlet
contribution to the polarization is now leading order, we need to be
concerned about the scale dependence discussed above.  Indeed, a NLO
calculation, in the modified double expansion scheme, is in
order. If, as in the case of the polarized rate, the octet
dominates, then we would expect only a slight longitudinal
polarization. Note that the NRQCD$_c$ polarization prediction has the
advantage that it only depends on one unknown matrix element, so once
the NLO calculation has been done, the prediction will be
comparatively robust.

\section{Conclusion}

There are several relevant questions to ask.  Is there any reason to
believe that there is any effective theory to correctly describe the
$\jp$?  We believe that the spin symmetry predictions for the ratio of
$\chi$ decays clearly answers this question in the affirmative.
Assuming that such an effective theory exists, then is it NRQCD$_c$ or
NRQCD$_b$?  As we have shown the two theories do indeed make quite
disparate predictions, which in principle should be easy to test.
However, these tests can be clouded by the issues of factorization and
the convergence of the perturbative expansion.

One would be justified to worry about the breakdown of factorization
in hadro-production at small transverse momentum. Indeed, in NRQCD$_c$
where the time scale for quarkonia formation is assumed to be the same
as the time scale for the hadronization of the remnants, it seems
quite likely that there could be order one corrections to
factorization.\footnote{This may be true as well in $B$
decays. However, since most of the time the $\jp$ will be going out
back to back with the remnants, one might expect factorization to be
more accurate.}  Thus any support, or lack thereof, for the theory
coming from these processes should be taken with a grain of salt. On
the other hand, for large transverse momentum one would expect
factorization to hold, with non-factorizable corrections suppressed by
powers of $m_c/p_T$.

As far as the perturbative expansion is concerned, it seems that for
most calculations the next-to-leading order results are indeed smaller
than the leading order result \cite{BR96,mp,NLO}.  However, the one
NNLO calculation performed, in the leptonic decay width \cite{BS}, is
not well behaved at this order, which is worrisome.  But, to
truly test the convergence of the expansion we should take ratios of
rates in order to eliminate the renormalon ambiguities \cite{BC}. When
this is done, it could very well be that the perturbative expansion is
well behaved. Until another rate is calculated at NNLO we will have to
be comforted by the fact that such cancellations have been seen to
occur explicitly in other heavy quark decays \cite{mren}.
 
With that said, let us gather the evidence in support of NRQCD$_c$ as
being the proper theory for the $\jp$. If one is willing to accept
that the extraction of the octet matrix elements from
CDF,\footnote{This extraction is not free of factorization issues
since the fit of the matrix elements involves use of data at rather
small values of transverse momentum. However, if a cut at $p_T=5{\rm\
GeV}$ is made, then the change in the fit is minimal.} then the fact
that the ratio is large for charmonia but seems to be small for bottomonia is
rather compelling. If once the statistics in the bottomonia sector
improve we find that there really is no hierarchy, then we believe
that this would be strong evidence for our hypothesis.
The fact that the $\jp$ radiative decay spectrum is
monotonically decreasing while one sees a bump at larger energy in
bottomonia also seems to lend credence to our hypothesis.  However,
the true litmus will come from the polarization measurements at large
$p_T$. The predictions of nearly 100\% polarization in NRQCD$_b$ is
quite robust. Whereas, in NRQCD$_c$ the polarization is diluted from
$M1$ transitions of the $O_8(^3S_1)$ operator.  Unfortunately, the
introduction of another unknown matrix element diminishes our
predictive power. However, this is not to say that we can not rule out
NRQCD$_c$. Indeed, NRQCD$_c$ also predicts a leveling off of the
polarization with positive $\alpha$.  So a measurement of longitudinal
(or, for $J/\psi$, zero) polarization would indeed negate our
hypothesis.

We would like to close with a caveat. In particular, it should be
pointed out that NRQCD$_c$ does not become exact in any
limit. Typically, in an effective field theory, we expect that the
ratio of sub-leading to leading contributions should vanish in a given
limit of QCD.  This gives us confidence that the theory MUST be
correct in asymptotia. Whether or not the real world leads to a well
behaved expansion though, is another question.  For NRQCD$_c$ we might
hope that as we take the limit $\lqcd/m\rightarrow 0$, we necessarily
get the correct answer. However, in this limit the soft modes become
perturbative and the power counting changes. That is, in this limit
the state becomes Coulombic and NRQCD$_b$ becomes the correct theory.
It may well be the case that in some observables the NRQCD$_c$
expansion is well behaved and in others it is not. Given that the
expansion parameter is around $1/3$, it seems reasonable to be
confident in those predictions for which the corrections are
suppressed by at {\it least} $\lqcd^2/m^2$ (modulo the convergence of
the perturbative expansion), as are the predictions discussed in this
paper.

\acknowledgements This work was supported in part by the
U.S. Department of Energy under grant numbers DOE-ER-40682-143 and
DE-AC02-76CH03000.
}


\begin{thebibliography}{99}
\bibitem{tab} These results reproduce those given in \cite{martin}
when $\lambda$ is taken to be 1 in this reference.
\end{thebibliography}
\end{document}